\documentclass{xpkas}

\def\year{2015} 
\def\month{January} 
\def\volume{00} 
\def\issue{0} 
\def\beginpage{1} 
\def\endpage{99} 
\def\received{October 31, 2014} 
\def\accepted{November 30, 2014} 
\date{Received \received ; accepted \accepted}

\title{Super-massive black hole mass scaling relations}

\author[1]{Alister W.~Graham}
\author[1,2]{Nicholas~Scott}
\author[3]{James M.~Schombert}  
\affil[1]{Centre for Astrophysics and Supercomputing, Swinburne University 
of Technology, Hawthorn, VIC, 3122, Australia; \email{AGraham@swin.edu.au}}
\affil[2]{Sydney Institute for Astronomy, School of Physics, University of
Sydney, NSW 2006, Australia;}
\affil[3]{Department of Physics, University of Oregon, Eugene, OR 97403, USA.}

\def\corrauthor{A. W. Graham}
\def\runningauthor{Graham, Scott \& Schombert}
\def\runningtitle{Super-massive black hole mass scaling relations}

\def\keywords{black hole physics; galaxies: bulges; galaxies: fundamental
  parameters; galaxies: nuclei}

\def\abstracttext{ 
Using black hole masses which span $10^5$--$10^{10} M_{\odot}$, the
distribution of galaxies in the (host spheroid stellar mass)--(black hole
mass) diagram is shown to be strongly bent.  While the core-S\'ersic galaxies
follow a near-linear relation, having a mean $M_{\rm bh}/M_{\rm sph}$ mass
ratio of $\sim$0.5\%, the S\'ersic galaxies follow a near-quadratic relation.  This
is not due to offset pseudobulges, but is instead an expected result arising
from the long-known bend in the $M_{\rm sph}$--$\sigma$ relation and a log-linear
$M_{\rm bh}$--$\sigma$ relation. 
}

\begin{document}
\pkashead 

\section{The $M_{\rm bh}$--$\sigma$ diagram}

\begin{figure}
\label{Fig1}
\centering
\includegraphics[width=70mm,angle=-90]{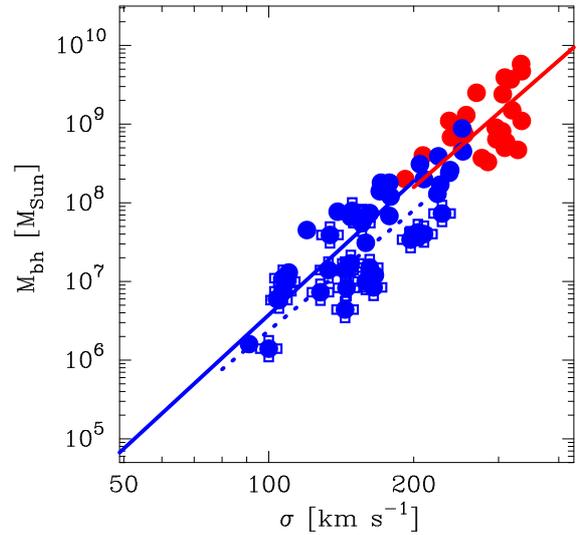}
\caption{Variation of the (black hole mass)--(velocity dispersion) figure~2a
  from Graham \& Scott (2013). 
  The red points are core-S\'ersic galaxies, while the blue points are
  S\'ersic galaxies. The crosses designate barred galaxies, which tend to be
  offset to higher velocity dispersions. The three lines are linear
  regressions, in which the barred S\'ersic galaxies and the non-barred
  S\'ersic galaxies are fit separately from the core-S\'ersic galaxies (none
  of which are barred).  
}
\end{figure}

As one of the most popular topics in astronomy over the past 15 years, the
$M_{\rm bh}$--$\sigma$ diagram (Ferrarese \& Merritt 2000; Gebhardt et
al.\ 2000) needs little introduction.  The relation between a galaxy's central
supermassive black hole mass and its velocity dispersion is shown in
Figure~\ref{Fig1} for 72 galaxies.  Taken from Graham \& Scott (2013, see also
McConnell \& Ma 2013), the non-barred S\'ersic galaxies can be seen to follow
the same $M_{\rm bh}\propto \sigma^{5.5}$ scaling relation as the non-barred
core-S\'ersic galaxies\footnote{To date, no barred core-S\'ersic galaxy is
  known.}.  The barred galaxies have a tendency to be offset to larger
velocity dispersions, a result explained in terms of elevated dynamics due to
the bar (Hartmann et al.\ 2013, see also Brown et al.\ 2013 and Debattista et
al.\ 2013).

In addition to bar dynamics, the observed velocity dispersion can be
overestimated due to strong rotational gradients within the inner region, as
was noted in Graham et al.\ (2011, their section~4.2.2 ).  Removing these
biases to obtain the spheroid's, rather than galaxy's, velocity dispersion
will reduce the observed velocity dispersion and likely decrease some of the
scatter in the $M_{\rm bh}$--$\sigma$ diagram\footnote{This challenge has been
  taken up by Woo (2015).}.  Most recently, from an expanded sample of 89
galaxies with directly measured black hole masses, Savorgnan \& Graham (2014)
find a slope of $\sim$6.34$\pm$0.80 for the 57 non-barred members, and a
vertical scatter of 0.53 dex in the $M_{\rm bh}$-direction.

\section{The $L_{\rm sph}$--$\sigma$ and $M_{\rm sph}$--$\sigma$ diagram}

\begin{figure*}[t]
\label{Fig2}
\centering
\includegraphics[angle=-90,width=70mm]{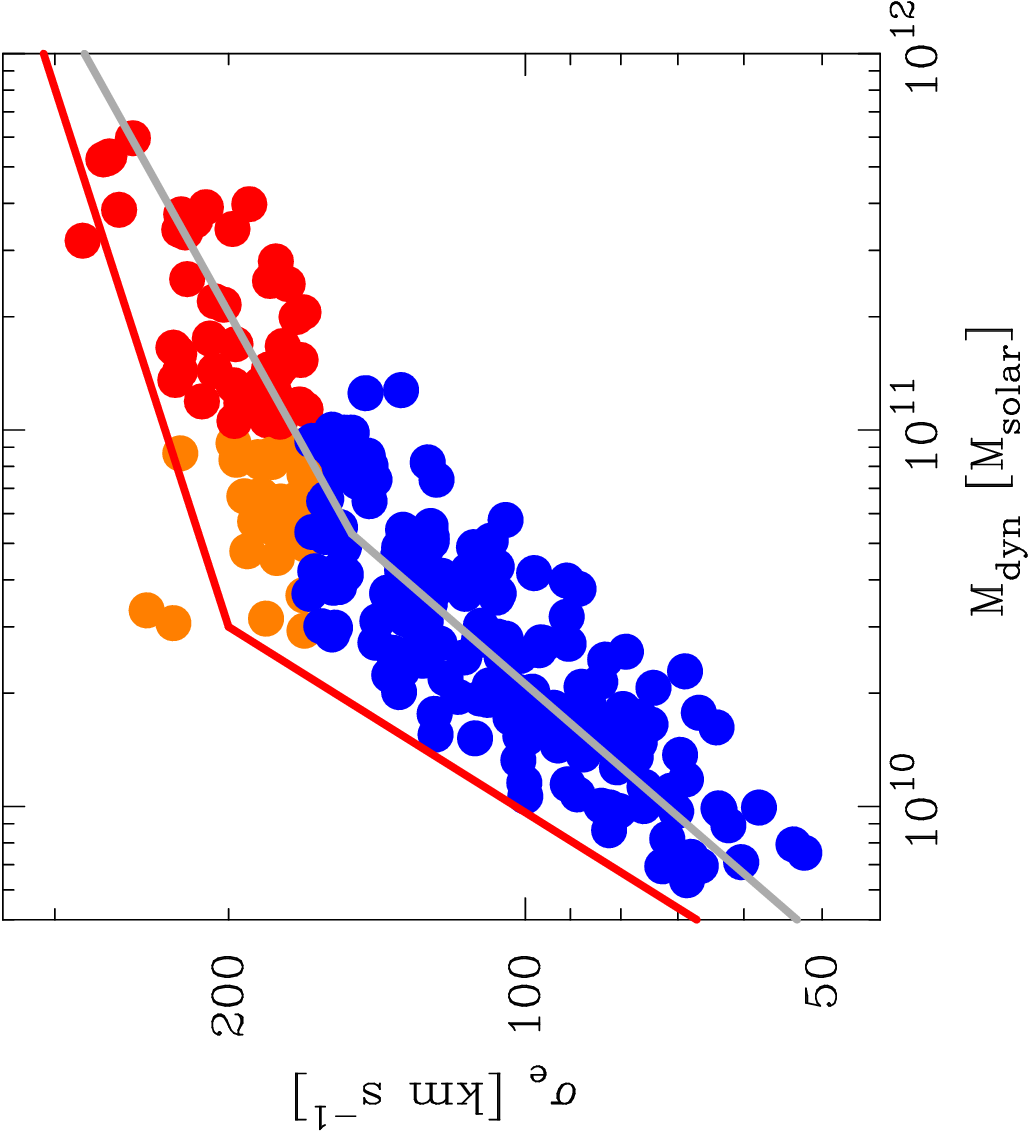} 
\hspace*{5mm}
\includegraphics[angle=-90,width=70mm]{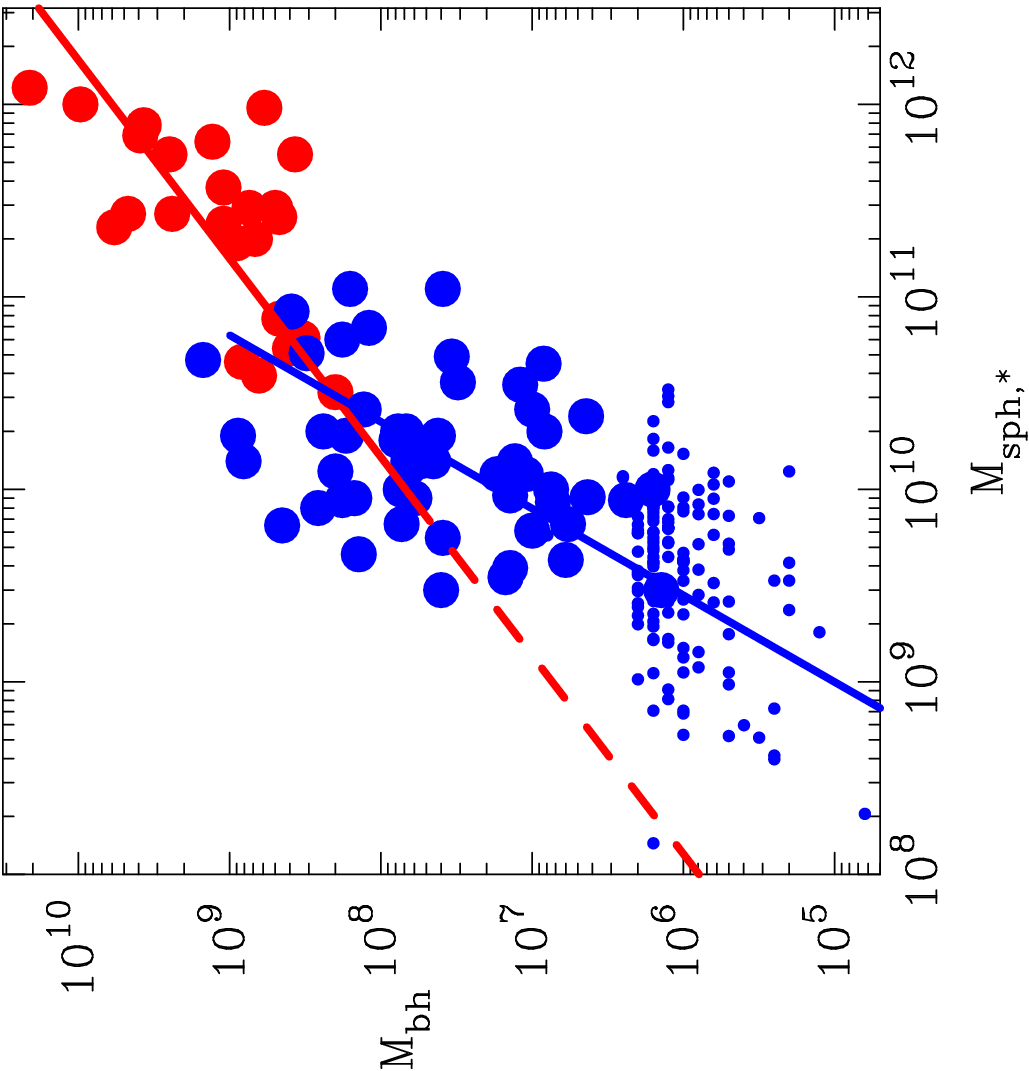}
\caption{
Left: 
  Adaption of figure~1 from Cappellari et al.\ (2013b), with the colour
  scheme roughly matching galaxy type such that orange indicates either a
  core-S\'ersic galaxy or a S\'ersic galaxy. The lines are from their figure. 
Right:
  Building on figure~3 from Scott et al.\ (2013), we have 
  added 139 AGN (smaller symbols) from Jiang et al.\ (2011a, b) --- see Graham \&
  Scott (2014) for details. 
  The red points are core-S\'ersic galaxies while the blue points are
  S\'ersic galaxies.  The regressional lines to the non-AGN data
  from Scott et al.\ (2013) reveal that the near-quadratic relation for the
  S\'ersic galaxies matches the AGN data well.  The systematic, rather than
  random, deviation from the near-linear core-S\'ersic relation is
  increasingly evident at lower masses. 
}
\vspace{5mm} 
\end{figure*}

Over half a century ago, Minkowski (1962) noted a correlation between
luminosity and velocity dispersion for early-type galaxies (see the review in
Section~3.3.3 of Graham 2013).  Schechter (1980) and Malumuth \& Kirshner
(1981) subsequently reported a slope of $\sim$5 for the luminous galaxies
(i.e.\ $L\propto \sigma^5$), and then Davies et al.\ (1983) reported a slope
of $\sim$2 for the low- and intermediate-luminosity early-type galaxies.
Samples containing mixtures of these two populations have slopes closer to 3
(e.g.\ Tonry et al.\ 1981) or 4 (e.g.\ Faber \& Jackson 1976) depending on the
relative number of bright to faint galaxies in one's sample.

The $L\propto \sigma^2$ relation extends from the lowest luminosity dwarf
elliptical galaxies ($\sigma \approx 20$ km s$^{-1}$, and stellar masses a few
times $10^8 M_{\odot}$) up to $M_{\rm sph,*}\sim$$10^{11} M_{\odot}$ upon
where massive spheroids and elliptical galaxies with partially depleted cores
dominate (e.g.\ Matkovi\'c \& Guzm\'an 2005; Evstigneeva et al.\ 2007; Forbes
et al.\ 2011; Kourkchi et al.\ 2012).  The massive galaxies follow the
relation $L\propto \sigma^{5–-6}$ (von der Linden et al.\ 2007; Liu et
al.\ 2008), with depleted cores starting to appear in galaxies with $\sigma
\gtrsim 170$ km s$^{-1}$ and becoming quite prevalent once $\sigma \gtrsim
230$ km s$^{-1}$ (e.g.\ Dullo \& Graham 2012).

In addition to the dwarf and intermediate luminosity early-type (S\'ersic)
galaxies ($-14 > M_B > -20.5$ mag) following the same log-linear $L$--$\sigma$
relation noted above, their unification as a single population is also evident
through the log-linear $L$--$n$ and $L$--$\mu_0$ relations that they share
(Young \& Currie 1994; Jerjen \& Binggeli 1998, see also Caon et al.\ 1993 and
Schombert 1986), where $n$ is the S\'ersic (1963) index and $\mu_0$ is the
central surface brightness. Furthermore, they display a similar behavior in
terms of the occurrence of a rotating stellar disk and various other kinematic
substructure (e.g.\ Emsellem et al.\ 2011; Scott et al.\ 2014; Toloba et
al.\ 2014).  Due to their systematically changing S\'ersic index with
luminosity (i.e.\ the $L$--$n$ relation), the difference between $\mu_0$ and
$\mu_e$ (the surface brightness at the effective half light radius $R_{\rm
  e}$) varies non-linearly with luminosity.  This produces the dramatically
curved $L$--$\mu_{\rm e}$ relation, and the curved $L$--$R_{\rm e}$ relation,
whose bright and faint arms have in the past been mis-interpreted as evidence
for a dichotomy between dwarf and intermediate luminosity early-type galaxies
because the curvature is greatest at $M_B \approx -18$ mag ($\approx
2$--$3\times10^{10} M_{\odot}$)\footnote{Due to the presence of disks in the
  lenticular galaxies, minor perturbations are expected and found (e.g.\ Janz
  \& Lisker 2008) about these unifying relations.}.  For those who are
interested to learn more about galaxy structure, Graham (2013) provides 
an historical and modern review with references to over 
500 papers, including many pioneer and often over-looked papers.

Naturally, the bent $L_{\rm gal}$--$\sigma$ relation mentioned above for
early-type galaxies maps into a bent $M_{\rm gal}$--$\sigma$ relation.  The
(dynamical mass)--(effective velocity dispersion) diagram from Cappellari et
al.\ (2013b; their figure~1) has been reproduced here in Figure~\ref{Fig2}a
using the same data\footnote{The velocity dispersions within $R_{\rm e}/8$
  from table~1 of Cappellari et al.\ (2013b) produce a somewhat similar
  distribution.} from table~1 of Cappellari et al.\ (2013a).  Their dynamical
mass is twice their Jeans Anisotropic Multi-Gaussian-Expansion (JAM) mass within the
effective half-light radius $R_{\rm e}$, and the `effective velocity 
dispersion' ($\sigma_{\rm e}$) is the velocity dispersion within $R_{\rm e}$.
%

At $\sigma_{\rm e} \sim 50$ km s$^{-1}$, the dynamical masses are around $4\times10^9
M_{\odot}$ (Figure~\ref{Fig2}a), while the spheroidal stellar masses are
around $10^9 M_{\odot}$ (Figures~\ref{Fig1} \& \ref{Fig2}b, assuming a common
black hole mass around $10^5 M_{\odot}$)\footnote{Note: For low mass spheroids
  the velocity dispersion profiles are rather flat, and $\sigma_{\rm e}
  \approx \sigma_{\rm e/8} \approx \sigma_0$.}.  To be consistent, this would
require early-type galaxies with $\sigma_{\rm e} \sim 50$ km s$^{-1}$ to have 3 times
as much dark matter as luminous matter, based on the JAM models.  This agrees
with an extrapolation of the data presented in figure~10 from Cappellari et
al.\ (2013b)\footnote{Some tension is noted with figure~14 from Forbes et
  al.\ (2008) which suggests that there may be roughly equal amounts, or
  little need for dark matter.}, and thus there is a consistency.

In passing we make two notes.  The bulges of spiral galaxies with 
$\sigma_{\rm e} \sim 50$ km s$^{-1}$ may not have such relatively high
dynamical-to-stellar masses if these galaxies' purported dark matter dominates
at larger radii.  Second, as one progresses from early-to-later type disk
galaxies, {\it on average} the radius $R_{\rm e,gal}$ will increasingly
resemble $R_{\rm e,disk}$ rather than $R_{\rm e,bulge}$ (i.e.\ $R_{\rm
  e,spheroid}$).  As such, use of virial mass estimators ($\sigma^2.R_{\rm
  e,gal}$) will increasingly over-estimate the dynamical mass of the bulge.

We write ``on average'' in the preceding paragraph because the spectrum of
disk galaxies, represented by the Hubble-Jeans tuning-fork sequence (Jeans
1919, 1928; Hubble 1926, 1936) or the ``Hubble comb'' for the Revised David
Dunlap Observatory system (van den Bergh 1976; Laurikainen et al.\ 2010, 2011;
Cappellari et al.\ 2011) may be better described by a kind of ``Hubble grid''
(Morgan \& Osterbrock 1969; Graham 2014) in which galaxies of each
morphological type (not just the S0 galaxies) span a range of bulge-to-disk
mass ratios.

\section{The $M_{\rm bh}$--$M_{\rm sph}$ diagram}

Graham \& Scott (2013) used $K_s$-band magnitudes from the Two Micron All-Sky
Survey (2MASS) Extended Source Catalogue (Jarrett et al.\ 2000) to confirm
that the $M_{\rm bh}$--$L_{\rm sph}$ relation is bent (Graham 2012).  Using
improved 2MASS magnitudes from the {\sc archangel} photometry pipeline
(Schombert \& Smith 2012), these revised spheroid magnitudes were converted
into stellar masses by Scott, Graham \& Schombert (2013), and the $M_{\rm
  bh}$--$M_{\rm sph}$ relation was shown to be bent.  Shown in
Figure~\ref{Fig2}b is the data from Scott et al.\ (2013) combined with data
for 139 Active Galactic Nuclei (AGN).  Virial estimates for the AGN black hole
masses are provided in Jiang et al.\ (2011a, b), and the apparent spheroid
magnitudes reported there have been converted into stellar masses by Graham \&
Scott (2014).  The lines shown in Figure~\ref{Fig2}b are from the fit in Scott
et al.\ (2013) to the core-S\'ersic and S\'ersic galaxies with directly
measured SMBH masses.  The core-S\'ersic relation has a slope of
0.97$\pm$0.14, see also Graham (2012) and Graham \& Scott (2013) which reports
that the mean $M_{\rm bh}/M_{\rm sph}$ ratio is 0.49\% for the core-S\'ersic
galaxies (see also Laor 2001).  The steeper S\'ersic $M_{\rm bh}$--$M_{\rm sph}$ relation
\begin{equation}
\log \frac{M_{\rm bh}}{M_{\odot}} = (2.22\pm0.58)\log\left[\frac{M_{\rm
      sph,*}}{2\times10^{10}M_{\odot}}\right] + (7.89\pm0.18)
\end{equation}
can be seen to match well with the distribution
of AGN data, and reveals that the S\'ersic $M_{\rm bh}$--$M_{\rm sph}$
relation extends down to $M_{\rm bh} \sim 10^5 M_{\odot}$.
This explains the steeper relations seen in the data of Laor (1998, 2001) and 
Wandel (2001). 

\section{Summary}

If $M_{\rm bh} \propto \sigma^{5.5}$, and $L \propto \sigma^2$ for S\'ersic
galaxies, then $M_{\rm bh} \propto L^{2.75}$.  Given $M_{\rm dyn}/L \propto
L^{1/3}$ (Cappellari et al.\ 2006), one has that $M_{\rm bh} \propto M_{\rm
  dyn}^{2.06}$ (or $M_{\rm bh} \propto M_{\rm dyn}^{2.44}$ if $M_{\rm bh}
\propto \sigma^{6.5}$, Savorgnan \& Graham 2014).  This bodes well with the
relation $M_{\rm bh} \propto M_{\rm sph,*}^{2.22\pm0.58}$ reported in Scott et
al.\ (2013).

If the sample of AGN from Jiang et al.\ (2011a,b) were associated with
pseudobulges having randomly low black hole masses relative to their host
bulge mass --- an idea originally proposed by Hu (2008) and Graham (2008) ---
then they would not display the distribution seen in Figure~\ref{Fig2}b.  They
would instead appear randomly offset to lower black hole masses rather than
following the near-quadratic $M_{\rm bh}$--$M_{\rm sph,*}$ relation.

This near-quadratic, or possibly super-quadratic, scaling relation 
has many implications.  For one, the
accretion and growth process, a popular topic at this meeting (e.g.\ Qiao
2015; Han 2015; Yang 2015, Taam 2015) does not obey a constant $M_{\rm
  bh}/M_{\rm sph}$ mass ratio.  There are also important implications for
gravitational radiation, another popular theme at this meeting (e.g.\ Hobbs
2015; Kang 2015; Kim 2015; Lee 2015).  Due to the relatively smaller black hole masses in
the lower-mass S\'ersic galaxies, which also typically house a dense nuclear
star cluster, the detectable number of low-frequency `extreme mass ratio
inspiral' events (Amaro-Seoane et al.\ 2014, and references therein) 
may be an order of magnitude lower than compared to
expectations if $M_{\rm bh}/M_{\rm sph} \approx 0.1$\% (Mapelli et al.\ 2012).
These and other consequnces of the new bent scaling relation 
are described in Graham \& Scott (2014).

\acknowledgments

This research was supported by Australian Research Council funding through
grants DP110103509 and FT110100263.

\end{document}